\documentclass[%
 aip,
 amsmath,amssymb,
 reprint,%
]{revtex4-1}

\usepackage{graphicx}
\usepackage{dcolumn}

\usepackage[utf8]{inputenc}
\usepackage[T1]{fontenc}
\usepackage{mathptmx}
\usepackage{etoolbox}

\makeatletter
\def\@email#1#2{%
 \endgroup
 \patchcmd{\titleblock@produce}
  {\frontmatter@RRAPformat}
  {\frontmatter@RRAPformat{\produce@RRAP{*#1\href{mailto:#2}{#2}}}\frontmatter@RRAPformat}
  {}{}
}%

\usepackage{siunitx}
\usepackage{upgreek}
\usepackage{xcolor}

\newcommand{\kB}{k_{\mathrm{B}}}

\newcommand{\St}{\mathrm{St}}
\newcommand{\Reynolds}{\mathrm{Re}}

\newcommand{\nuLB}{\nu_{\mathrm{LB}}}

\newcommand{\um}{\micro\metre}

\newcommand{\latin}[1]{{\itshape #1}}
\newcommand{\eg}{e.g.\ }

\newcommand{\etal}{\latin{et al.}}

\newcommand{\Eqref}[1]{Eq.~\eqref{#1}}

\newcommand{\Figref}[1]{Fig.~\ref{#1}}

\newcommand{\revs}[1]{{\color{black} #1}}
\newcommand{\revstwo}[1]{{\color{black} #1}}

\usepackage[colorlinks,citecolor=red,urlcolor=blue,bookmarks=false,hypertexnames=true]{hyperref} 

\makeatother
\begin{document}

\title[Filtration efficiency of a woven fabric]{Modelling the filtration efficiency of a woven fabric: The role of multiple lengthscales}

\author{Ioatzin Rios de Anda}
\affiliation{H.\ H.\ Wills Physics Laboratory, University of Bristol, Bristol BS8 1TL, United Kingdom}
\affiliation{School of Mathematics, University Walk, University of Bristol, BS8 1TW, United Kingdom}
\author{Jake W. Wilkins}
\affiliation{Department of Physics, University of Surrey, Guildford, GU2 7XH, United Kingdom}
\author{Joshua F. Robinson}
\affiliation{H.\ H.\ Wills Physics Laboratory, University of Bristol, Bristol BS8 1TL, United Kingdom}
\affiliation{Institut f\"ur Physik, Johannes Gutenberg-Universit\"at Mainz, Staudingerweg 7-9, 55128 Mainz, Germany}
\author{C. Patrick Royall}
\affiliation{Gulliver UMR CNRS 7083, ESPCI Paris, Universit\'{e} PSL, 75005 Paris, France}
\author{Richard P. Sear}
\affiliation{Department of Physics, University of Surrey, Guildford, GU2 7XH, United Kingdom}
\email{r.sear@surrey.ac.uk}
 \homepage{https://richardsear.me/}

\begin{abstract}
During the COVID-19 pandemic, many millions have worn masks made of woven fabric, to reduce the risk of transmission of COVID-19. Masks are essentially air filters worn on the face, that should filter out as many of the dangerous particles as possible. Here the dangerous particles are the droplets containing virus that are exhaled by an infected person. Woven fabric is unlike the material used in standard air filters. Woven fabric consists of fibres twisted together into yarns that are then woven into fabric. There are therefore two lengthscales: the diameters of: (i) the fibre and (ii) the yarn. Standard air filters have only (i). To understand how woven fabrics filter, we have used confocal microscopy to take three dimensional images of woven fabric. We then used the image to perform Lattice Boltzmann simulations of the air flow through fabric. With this flow field we calculated the filtration efficiency for particles \revs{a micrometre and larger in diameter}. \revs{In agreement with experimental measurements by others, we} find that for particles in this size range, filtration efficiency is low. \revs{For particles with a diameter of 1.5 micrometres our estimated efficiency is in the range 2.5 to 10\%}. The low efficiency is due to most of the air flow being channelled through relatively large (tens of micrometres across) inter-yarn pores. So we conclude that fabric is expected to filter poorly due to the hierarchical structure of woven fabrics.
\end{abstract}

\maketitle

\section{Introduction}

During the COVID-19 pandemic, billions of people have worn masks (face coverings) to protect both themselves and others from infection \cite{greenhalgh2020,greenhalgh2021,stadnytskyi2021,wang2021,xu2020}. There are three basic types of mask or face covering. Surgical masks and respirators are made of non-woven materials, while cloth masks are made of woven material. Filtration of air by non-woven materials is well studied \cite{wang2013}. However, pre-pandemic, very little research was done into filtration by woven materials, which have a different structure to that of non-woven materials. Here we try and address this, by studying how a woven fabric filters small particles out of the air.

\begin{figure}[htb!]
  \begin{center}
      \includegraphics[width=7.5cm]{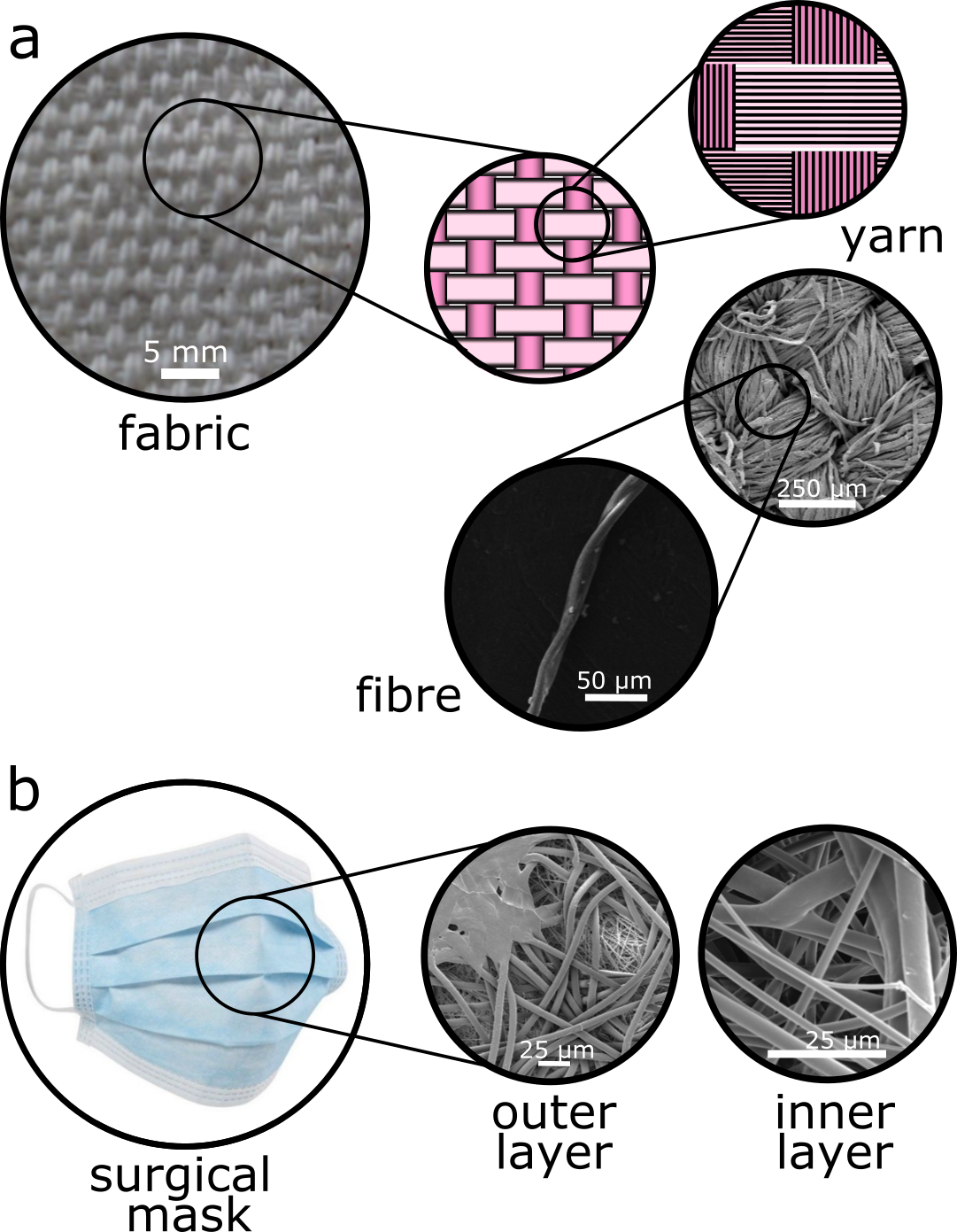}
      \caption{\revs{(a)} Fabric is a porous material with structure on multiple lengthscales. \revs{For the top three images,} from left to right we look at successively smaller lengthscales. At the largest lengthscale, fabric is a lattice woven from perpendicular yarns that go over and under other yarns at right angles to them. In the \revs{middle} schematic, vertical yarns are shown as dark pink, horizontal yarns as pale pink. As illustrated in both the \revs{top right} schematic and the SEM images on the right, these yarns are made by twisting together many much smaller fibres. At the \revs{bottom of (a)} we show a single fibre. Fibres are of order $\SI{10}{\micro\metre}$ \revs{in diameter} while yarns are a few hundred $\SI{}{\micro\metre}$ across. \revs{(b) From left to right we have an image of a typical surgical mask, and SEM images of the fibres of which it is made. Note that the fibres are randomly distributed, there is no lengthscale above that of the fibres, and the fibres in a filtering inner layer of a surgical masks typically have diameters a little less than $\SI{10}{\micro\metre}$, Lee \etal\cite{lee2021b} quote a mean diameter of $\SI{5.5}{\micro\metre}$.}}
      \label{fig:schem}
  \end{center}
\end{figure}

Woven fabrics have a very different structure from surgical masks.  \revs{We compare the structures of woven fabrics and surgical masks in \Figref{fig:schem}}. Surgical masks are meshes of long thin fibres \cite{wang2013}, of order ten micrometres thick, \revs{see \Figref{fig:schem}(b)}. \revs{However, fabrics are different, they} are woven from cotton (or polyester, silk, $\ldots$) yarn. Cotton yarn is a few hundred micrometres thick, and is composed of cotton fibres each of order ten micrometres thick. These fibres are twisted into yarns, which are in turn woven into the fabric \cite{warren2018}, see \Figref{fig:schem}. This two-lengthscale (fibre and yarn) hierarchical structure of fabric is known to affect the fluid flow through fabric, because it has been studied in the context of laundry \cite{vandenbrekel1987,shin2018}. However, there has been little effort to study its effect in the context of particle filtration \cite{robinson2021}.

To understand how woven fabrics filter air, we started by using a confocal microscope to obtain a three-dimensional image of a sample of fabric, at a spatial sampling rate of $\SI{1.8}{\micro\metre}$. This image is then used as input to Lattice Boltzmann simulations of air flow inside a woven face mask during breathing. That flow field is then used to calculate large numbers of particle trajectories through the fabric to estimate filtration efficiencies.

\subsection{Previous work on filtration by woven fabrics}

Konda \etal\cite{konda2020,konda2020b}, Duncan \etal\cite{duncan2020} and Sankhyan \etal\cite{sankhyan2021}, have all measured filtration efficiencies for a number of fabrics. They all studied the filtration of particles in the size range we consider, which is $\ge\SI{1}{\micro\metre}$.  Zangmeister et al.\cite{zangmeister2020} has studied filtration for smaller particles. Note that the original measurements of Konda and coworkers suffered from methodological problems\cite{rule2020,carr2020,hancock2020,konda2020b}, which were later corrected \cite{konda2020b}.

\revs{This work directly measured filtration efficiencies but did not image the fabric in three dimensions. Lee \etal\cite{lee2021}, Du \etal\cite{du2021}, and Lee \etal\cite{lee2021b} have all imaged the filtration media of surgical masks \cite{lee2021,du2021}, or of respirators with the surface charges removed making the filtration media similar to that of many surgical masks \cite{lee2021b}. However, Lee \etal\cite{lee2021} and Du \etal\cite{du2021} did not use this imaging data to compute filtration efficiencies, while Lee \etal\cite{lee2021b} only performed relatively limited studies of filtration efficiency.}

\subsection{Evidence that droplets approximately a micrometre in diameter carry infectious SARS-CoV-2 virus}

The literature on COVID-19 transmission is large but it is worthwhile briefly summarising the part most relevant to this work. The breath we exhale is an aerosol of small mucus droplets in air that is warm and humid because it has come from our lungs \cite{bourouiba2020}. These droplets range in size from much less than a micrometre to hundreds of micrometres \cite{johnson2011}. Vocalisation (i.e. speech or singing) produces more aerosol than ordinary breathing \cite{johnson2011,gregson2021,asadi2019}. The peak in the size distribution function of exhaled droplets is around $\SI{1.6}{\micro\metre}$ --- this is the count median diameter of Johnson and coworkers \cite{johnson2011}. 

The median diameter of $\SI{1.6}{\micro\metre}$ is for droplets as exhaled in our breath, \revstwo{breath} which is essentially saturated with water vapour, \revstwo{i.e., at essentially 100\% relative humidity (RH) \cite{bourouiba2020}.}
\revstwo{It takes only a few milliseconds for droplets to pass through a mask filter (see section \ref{sec:diffusion}) and this short time combined with the 100\% RH, means that droplets do not evaporate while passing out through a mask filter.}. \revstwo{If a person inhales another person's breath more-or-less directly, for example if they are close and talking to each other, then the droplets inhaled will not have left the humid breath, and still have the same diameter as when they were exhaled.}

\revstwo{However, when our breath mixes with room air \cite{bourouiba2020,abkarian2020,bourrianne2021,bourrianne2021b},} the humidity drops. \revstwo{Then micrometre-sized} droplets evaporate in \revstwo{of order $\SI{10}{\milli\second}$} \cite{netz2020}. After this evaporation, the droplet diameter \revs{is smaller by a factor of two to three} \cite{netz2020,johnson2011,robinsonPanic2020}. So, typical droplet sizes are around $\SI{1.6}{\micro\metre}$ as we breath them out through a mask, but around \revs{$0.5$ to} $\SI{0.8}{\micro\metre}$ when we breath them in. \revstwo{We do not expect droplets to pick up significant amounts of water on inhalation through a filter, as the droplets will be in air from the surroundings, and they spend only a few milliseconds passing through the filter.}

\revstwo{Both $\SI{1.6}{\micro\metre}$, and around $0.5$ to $\SI{0.8}{\micro\metre}$} are approximate \revs{(count)} medians of broad distributions \cite{johnson2011}. \revs{Due to this evaporation after exhalation, there are two sets of droplet size distributions to consider when studying filtration, with the distribution on exhalation being two to three times larger in diameter than on inhalation. The particles that need to be filtered for source control are larger than need to be filtered to protect the wearer.}

Coleman and coworkers \cite{coleman2021} found SARS-CoV-2 viral RNA in both particles with diameters smaller than and larger than $\SI{5}{\micro\metre}$, and found that most of the viral RNA was in droplets with diameters less than $\SI{5}{\micro\metre}$. These diameters are after evaporation. Santarpia and coworkers \cite{santarpia2021} found infectious virus in particles both with diameters $<\SI{1}{\micro\metre}$ and in the range $1$ to $\SI{4}{\micro\metre}$, but not in particles larger than $\SI{4.1}{\micro\metre}$. Hawks and coworkers \cite{hawks2021} were also able to obtain infectious virus in aerosols smaller than $\SI{8}{\micro\metre}$. It should be noted that the study of Hawks and coworkers was of infected hamsters not humans. Finally, Dabisch and coworkers infected macaques with an aerosol of droplets with median diameter $\SI{1.4}{\micro\metre}$ \cite{dabisch2021}.  This body of very recent work suggests that aerosol particles of order a micrometre carry most of the virus.

It is also worth noting that Coleman and coworkers \cite{coleman2021} also found that amount of viral RNA varied widely from one person to another. Some infected people breathed out no measurable RNA. Those that did breathed out an amount that varied by a factor of almost a hundred. Viral RNA was found even for those who never developed COVID-19 symptoms, i.e., who always remained asymptomatic. 

As we state above, we use \lq droplet\rq~ to cover all sizes from much less than a micrometre to hundreds of micrometres and more. This is in line with the aerosol and fluid mechanics literature, but some work in the medical literature reserves \lq droplet\rq~for diameters over $\SI{5}{\micro\metre}$, despite there being no justification for this distinction \cite{tang2021,randall2021}.

\subsection{Evidence that masks filter out SARS-CoV-2}

Adenaiye and coworkers \cite{adenaiye2021} studied the effect of masks on the amount of viral SARS-CoV-2 RNA breathed out. This study tested a wide range of masks as participants were asked to bring their own masks. They found that in \lq fine aerosols ($<\SI{5}{\micro\metre}$)\rq, the masks reduced the amount of viral RNA detected by 48\% (95\% confidence interval 3 to 72 \%), while for larger aerosols, masks reduced the viral RNA by 77\% (95\% confidence interval 51 to 89 \%). Here, $\SI{5}{\micro\metre}$ is presumably the evaporated diameter (not radius) but this was not specified by the authors.

\subsection{Mechanism of filtration}

Filtration is traditionally ascribed to a sum of four mechanisms \cite{wang2013}. The idea being that a particle with zero size, zero inertia, zero diffusion, and zero charge, will follow the streamlines perfectly and not be filtered out. However, deviations from any one of those four conditions can cause a collision and hence filtration. 

The four mechanisms are:
\begin{enumerate}
    \item \emph{Interception:} Particles whose centre of mass follows streamlines perfectly can still collide with fibres, if the particles have a \revs{non-zero} size. This is a purely geometric mechanism, that does not require inertia. 
    \item \emph{Inertial:} With inertia, particles cannot follow the air streamlines perfectly. Whereas a streamline goes around an obstacle, a particle with inertia will deviate from the streamline and so may collide.
    \item \emph{Diffusion:} Particles diffuse in air, creating further deviations from streamlines and thus potential collisions with the obstacle.
    \item \emph{Electrostatic interactions:} Charges, dipole moments etc, on the fibres and on the droplets will interact with each other. If they pull the two towards each other, this will enhance filtration. Cotton fibres have no charge distribution as far as we know, so we do not expect this to be a significant mechanism here.
\end{enumerate}
Note that in practice these mechanisms are never completely independent \cite{wang2013}.

Flow through masks is sufficiently slow, and the lengthscales are sufficiently small, that the flow is close to Stokes flow, i.e., the Reynolds number is small. This means that streamlines do not depend on flow speed/pressure difference. In turn, this implies that interception filtration is independent of flow speed. Inertial filtration becomes more important with increasing flow speed, as the faster moving particles have more inertia. While diffusion filtration becomes less efficient at faster flow speeds, as then particles spend shorter times passing through the mask. The particles then have less time to diffuse into the material of the mask, and be filtered out.

Here, we will focus on particles a micrometre and larger, where diffusion is less important as a filtration mechanism because particles this large diffuse slowly. So we will focus on interception and inertial filtration. However, in the conclusion we will return to filtration by diffusion and argue that filtration by diffusion in our fabric should be very inefficient.

The remainder of this paper is laid out as follows. The next section describes how we imaged the fabric and analysed the imaging data. The third section describes our Lattice Boltzmann (LB) simulations of air flow through the mask. Then the next section characterises this air flow. The fifth and sixth sections have our method for calculating particle trajectories and our results for filtration, respectively. The seventh section briefly discusses filtration via diffusion. The last section is a conclusion.

\begin{table}[b!]
  \begin{center}
  \begin{ruledtabular}
  \begin{tabular}{ccc}
    Area of sample (\SI{}{\centi\metre\squared}) & mass (\SI{}{\gram}) & mass/area (\SI{}{\gram\per\centi\metre\squared})\\
    \hline
   1 &	0.01210 	& 0.01210 \\
   2.25 &	0.02742 &	0.01219 \\
   4 &	0.04813 &	0.01203 \\
  \end{tabular}
  \end{ruledtabular}
  \caption{
    Table of measurements of the mass of samples of the fabric, used to determine its mass per unit area.
  }
  \label{tab:mass_density}
  \end{center}
\end{table}

\section{Image acquisition and analysis of a sample of woven fabric}

\revs{In order to study filtration by woven fabric, a high-resolution 3D image of the fabric is needed. We used confocal optical imaging to obtain an image of fabric, at voxel size of $\SI{1.8}{\micro\metre}$. Recent work by Lee \etal\cite{lee2021} and by Du \etal\cite{du2021} has used X-ray tomography to obtain 3D images of the internal structure of surgical masks, but, to our knowledge, nobody has been able to image woven fabric or to use confocal microscopy for this purpose, before.}

The fabric was obtained from a commercial fabric mask. Square pieces of 1, 2.25 and $\SI{4}{\centi\metre\squared}$ were weighed individually, giving a
mass per unit area of $\SI{120}{\gram\per\metre\squared}$, see Table \ref{tab:mass_density}.
Using brightfield optical microscopy (Leica DMI3000 B) with a Leica 4x objective, we estimated the thickness of the fabric in air to be $285\pm\SI{24}{\micro\metre}$, which we determined through different measurements along the fabric. Using the mass density of cotton, $\rho_c$, from Table \ref{tab:numbers}, this corresponds to the fabric being on average about 28\% cotton fibres and 72\% air.

\subsection{Image acquisition}

In order to study the 3D structure of the fabric, square pieces of $\SI{0.5}{\centi\metre}$ of cotton were dyed with fluorescein (Sigma Aldrich) following Baatout \etal\cite{baatout2019}. The dyed cotton squares were then washed in deionised water to eliminate any dye excess and left to dry under ambient conditions for 48~hours. Once dried, the fabric was re-submerged in 1,2,3,4-tetrahydronaphthalene (tetralin, Sigma Aldrich). We chose this solvent due to its refractive index being close to the index of cotton ($\eta_{D\mathrm{tetralin}} = 1.544$\cite{gong2012} and $\eta_{D\mathrm{cotton}} =1.56-1.59$\cite{morton2008}). Such matching is needed to allow imaging with fluorescence confocal microscopy.

The dyed fabric samples were immersed in tetralin. They were confined in cells constructed using three coverslips on a microscope slide. Two of the coverslips acted as a spacer, and they were sealed using epoxy glue. The spacing coverslips have a height of $\SI{0.56}{\milli\metre}$, which prevented fabric compression. A confocal laser scanning microscope Leica TCS SP8 equipped with a white light laser, was used to study the fibre structures, using a Leica  HC PL APO 20x glycerol immersion objective with a 0.75 numerical aperture and a correction ring. The excitation/emission settings used for the fluorescein dye were 488 and \SI{500}{\nano\metre}, respectively. Scans of the cell in the $z$-axis were acquired to analyse the fibre network in 3D, where care was taken to ensure the pixel size ($\SI{1.8}{\micro\metre}$) was equal along all axes.

\begin{figure}[htb!]
  \begin{center}
    \includegraphics[width=7.5cm]{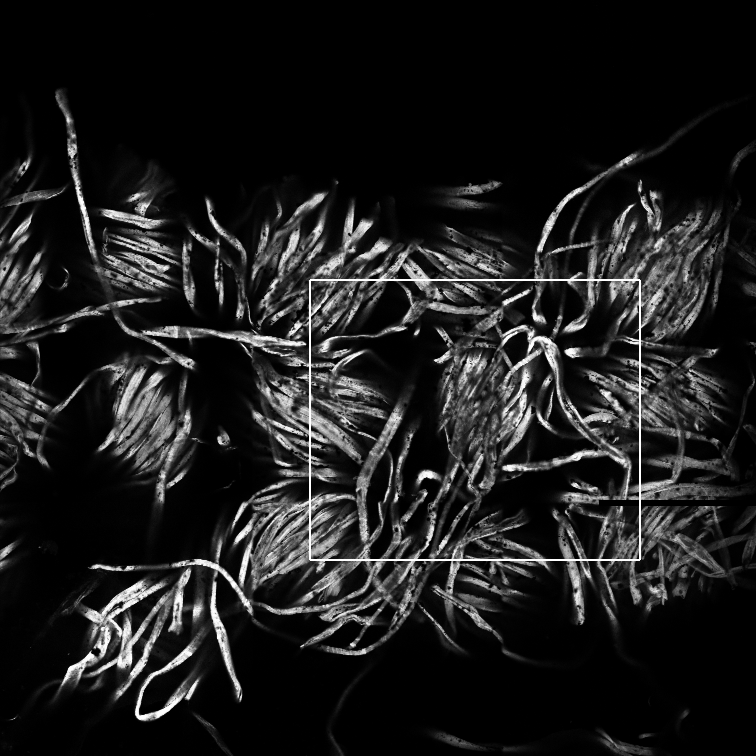}
      \caption{Slice (number 19, starting at 0) of the confocal image of the fabric. Slice is in the $xy$ plane. The area simulated using LB is enclosed by a white box.}
      \label{fig:Ioatzin_image}
  \end{center}
\end{figure}

The confocal microscopy data is in the form of a stack of $n_z=62$ images of the $xy$ plane, each of which is $n_x=756$ by $n_y=756$ voxels. Each voxel is a cube of side $\SI{1.8}{\micro\metre}$, see Table \ref{tab:fabric}. Slice number 19 (starting at zero) is shown in \Figref{fig:Ioatzin_image}. In each slice, approximately two-thirds of the field of view is taken up with a strip of the fabric, which runs left to right in \Figref{fig:Ioatzin_image}.

\begin{table}[tbh!]
  \begin{center}
  \begin{ruledtabular}
  \begin{tabular}{ccc}
    Quantity & Value & Reference \\
    \hline
    \multicolumn{3}{c}{
    \textbf{Air}} \\
    \hline
    mass density & \SI{1.2}{\kilogram\per\metre\cubed} & \onlinecite{crc} \\
    dynamic viscosity $\mu$ & \SI{1.8e-5}{\pascal\second} & \onlinecite{crc} \\
    kinematic viscosity $\nu$ & \SI{1.5e-5}{\metre\squared\per\second} & \onlinecite{crc} \\
    \hline
    \multicolumn{3}{c}{
    \textbf{Water/mucus}} \\
    \hline
    mass density $\rho_p$ (water) & \SI{998}{\kilogram\per\metre} & \onlinecite{crc}\\
    dynamic viscosity (mucus) & \SI{0.1}{\pascal\second} & \onlinecite{gittings2015} \\
    mucus/air surface tension
    $\gamma$  & $\SI{0.05}{\newton\per\metre}$ & \onlinecite{gittings2015} \\
    \hline
     \multicolumn{3}{c}{
    \textbf{Cotton fibres}} \\
    \hline
    mass density $\rho_c$  &   $\SI{1500}{\kilo\gram\per\metre\cubed}$ & \onlinecite{gordon2006}\\
    \hline
    \multicolumn{3}{c}{\textbf{Typical breathing flow rates}} \\
    \hline
    tidal breathing at rest
    & \SI{6}{\litre\per\minute} & \onlinecite{caretti2004} \\
    during mild exertion
    & \SI{20}{\litre\per\minute} & \onlinecite{caretti2004} \\
    during moderate exertion
    & \SI{30}{\litre\per\minute} & \onlinecite{caretti2004} \\
    during maximal exertion
    & \SI{85}{\litre\per\minute} & \onlinecite{caretti2004} \\
    \hline
    \multicolumn{3}{c}{\textbf{Average flow speeds}} \\
    \hline
    effective mask area & \SI{190}{\centi\metre\squared} & \onlinecite{coffey1999} \\
    flow speed (rest)
    & \SI{0.5}{\centi\metre\per\second} \\
    flow speed (mild)
    & \SI{1.8}{\centi\metre\per\second} \\
    flow speed (moderate)
    & \SI{2.7}{\centi\metre\per\second} \\
    flow speed (maximal)
    & \SI{7.5}{\centi\metre\per\second} \\
  \end{tabular}
  \end{ruledtabular}
  \caption{
    Table of parameter values for masks, air, water and mucus; all at \SI{20}{\celsius} and atmospheric pressure \SI{e5}{\pascal}.
    Note that small droplets dry rapidly and this will cause their viscosity to increase.
    Flow rates are determined from the volume typically exhaled during one minute.
    Moderate exertion is defined as that readily able to be sustained daily during 8 hours of work, whereas maximal exertion is the upper limit of what can be sustained for short periods of time (\eg during competitive sports).
    Flow speeds are calculated for the stated mask area and flow rates assuming perfect face seal.
  }
  \label{tab:numbers}
  \end{center}
\end{table}

Of the 62 slices, image quality in the bottom ten is poor, due to attenuation from the imperfect refractive index matching. So in effect, we can obtain good images for 52 slices, 
i.e.\ we can reliably image a section of fabric that is approximately $\SI{93.6}{\micro\metre}$ thick.

\subsection{Fibre size distribution}

\begin{figure}[htb!]
  \begin{center}
    \includegraphics[width=7.5cm]{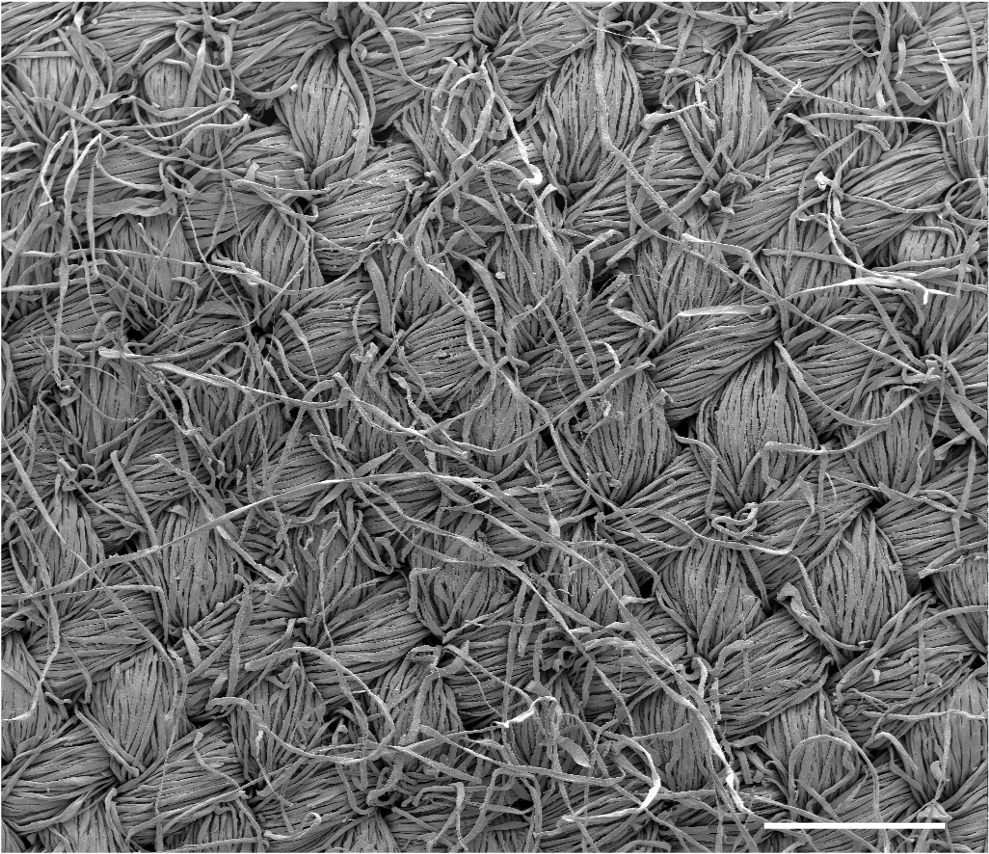}
      \caption{A Scanning Electron Microscope (SEM) image of the surface of our fabric. The fabric has been coated with gold/palladium. Secondary electron images were taken at 8 kV with a 100x magnification. Scale bar = $\SI{500}{\micro\metre}$.}
      \label{fig:SEM}
  \end{center}
\end{figure}

To obtain estimates of the distribution of fibre diameters we imaged the surface of the fabric using a scanning electron microscope (FEI Quanta 200 FEGSEM, Thermo Fisher Scientific), see \Figref{fig:SEM}. We then estimated the diameter of at least 50 fibres from this image, and obtained the mean and standard deviation of fibre diameters as
$16.7\pm \SI{4.8}{{\micro\metre}}$,which we determined by analysing SEM images.

\subsection{Image analysis}

The analysis of the image stack output by the confocal microscope was performed in Python using the
the OpenCV \cite{opencv_library} and cc3d \cite{cc3d} packages.
The confocal image stack is processed as follows:
\begin{enumerate}
\item We first delete the fibre voxels in the bottom ten slices due to the poorer image quality, leaving us with 52 slices of imaged fabric. We then add 200 slices to the top, and 200 slices to the bottom, all of each are of entirely zero intensity voxels. These additional slices are needed as the array produced for the simulations needs to cover fluid flow into and out of the fabric, i.e., we cannot just simulate flow inside the fabric, we need the approach and exit flows.
\item We then blur the image by convolving with a three-dimensional
Gaussian filter that is implemented as a sequence of 1-D convolution filters, with a standard deviation $\sigma_{B}=1$ voxel side ($\SI{1.8}{\micro\metre}$).
\item Next we threshold the blurred image, setting all voxels with values less than the threshold value $T=10$ to zero, and all voxels greater than or equal to the threshold value to one. Thus we get a binary image.
\item Then we use a 3D connected components algorithm to identify the connectivity of voxels that are one. We assign each voxel with value one to a cluster of connected voxels. All voxels of value one that are part of clusters of size $N_{CL}=25$ or less are set to zero, all other voxels of value one, are assumed to be fibre voxels. N.B. Applying the Gaussian filter greatly reduces the number of connected clusters we obtain.
\end{enumerate}

It is worth noting that step four only deletes a total of  507  voxels
while keeping 11681929 voxels so deleting a few isolated clusters has very little effect, and that in the final array almost 99.9\% of the voxels are part of the largest cluster. \revs{This is as we should expect. Most voxels should be in a single cluster, as the fabric needs to be one connected structure in order not to fall apart \cite{warren2018}. Varying the width of the Gaussian filter in the range $0.5$ to $2$ voxels has little effect. The number of voxels deleted does increase as $\sigma$ decreases, but at $\sigma=0.5$ (and a threshold $T=10$) we still only delete 3099 voxels from over eleven million, and the largest cluster has over 99.8\% of the voxels.}

\revs{Varying the threshold $T$ (keeping $\sigma=1$) in the range $T=5$ to 15, varies the number of fibre voxels by of order 10\%, from 13.6 million for $T=5$ to 10.2 million for $T=15$. Reducing the value of $T$ makes the fibres and yarns thicker and so the gaps between narrower. This suggests that there is an uncertainty of about 10\% in the volume of our fibres and yarns. Finally, varying the minimum cluster size $N_{CL}$ has little effect. Increasing it from 25 to 50 only increases the total number of fibre voxels deleted from 507 to 922, out of over 11 million (at $T=10 $ and $\sigma=1$).}

\begin{figure}[htb!]
  \begin{center}
    (a)\includegraphics[width=6.5cm]{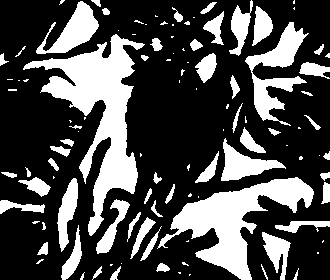}\\
    (b)\includegraphics[width=6.5cm]{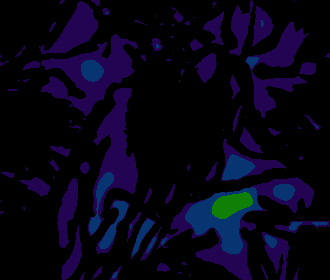}
      \caption{(a) The thresholded and so binary image produced by image analysis of the area in the white box in \Figref{fig:Ioatzin_image}. Fibre voxels are in black and air voxels are in white.
      (b) Heatmap of the $z$ component of the velocity in the same area. Again black is the fabric. Dark purple, blue and pale green are velocities less than the mean, between the mean and ten times the mean, and over ten times the mean velocity, respectively. Both images are $\SI{594}{\micro\metre}$ $\times$ $\SI{504}{\micro\metre}$. }
      \label{fig:threshold+flow}
  \end{center}
\end{figure}

\begin{table}[tbh!]
  \begin{center}
  \begin{ruledtabular}
  \begin{tabular}{cc}
    Quantity & Value \\
    \hline
       \multicolumn{2}{c}{
    \textbf{Fabric imaged}} \\
    \hline
    cubic voxel side length & \SI{1.8}{\micro\metre}\\
    total thickness imaged & 62 voxels = \SI{111.6}{\micro\metre} \\
    thickness used & $L_F=52$ voxels = \SI{93.6}{\micro\metre}\\
    area imaged & 756 $\times$ 756 voxels \\
         & = $\SI{1360.8}{\micro\metre}$ $\times$ $\SI{1360.8}{\micro\metre}$\\
             area used & $n_x=310$ to $310+330$ \\
              & $n_y=280$ to $280+280$ \\
         & = $\SI{594}{\micro\metre}$ $\times$ $\SI{504}{\micro\metre}$\\
    yarn lattice constants & $\SI{297}{\micro\metre}$ and $\SI{252}{\micro\metre}$  \\
    Threads per inch (TPI) & 186 \\
    \hline
    \multicolumn{2}{c}{
    \textbf{Lattice Boltzmann parameters}} \\
    \hline
    box size $n_x\times n_y\times n_z$ & $330\times 280 \times 462$\\
            & = $\SI{594}{\micro\metre}$ $\times$ $\SI{504}{\micro\metre}$ $\times$ $\SI{471.6}{\micro\metre}$\\
    Darcy velocity $U=Q/A$ & $5.6\times 10^{-7}$\\
    Re for lengthscale $\SI{297}{\micro\metre}$ & $6\times 10^{-4}$ \\
    pressure drop  & $6.7\times 10^{-6}$ \\
  \end{tabular}
  \end{ruledtabular}
  \caption{
    Table of parameter values for the fabric we have imaged, and for
    our Lattice Boltzmann simulations. TPI is calculated by adding together number of yarns per inch along $x$ and along $y$.
  }
  \label{tab:fabric}
  \end{center}
\end{table}

\subsection{Region of the fabric studied}

The fabric is essentially a rectangular lattice, woven from yarns that cross at right angles. Estimated lattice constants are in Table \ref{tab:fabric}. The lattice constants are around 20 times the average fibre diameter.

\begin{figure}[htb!]
  \begin{center}
    \includegraphics[width=7.5cm]{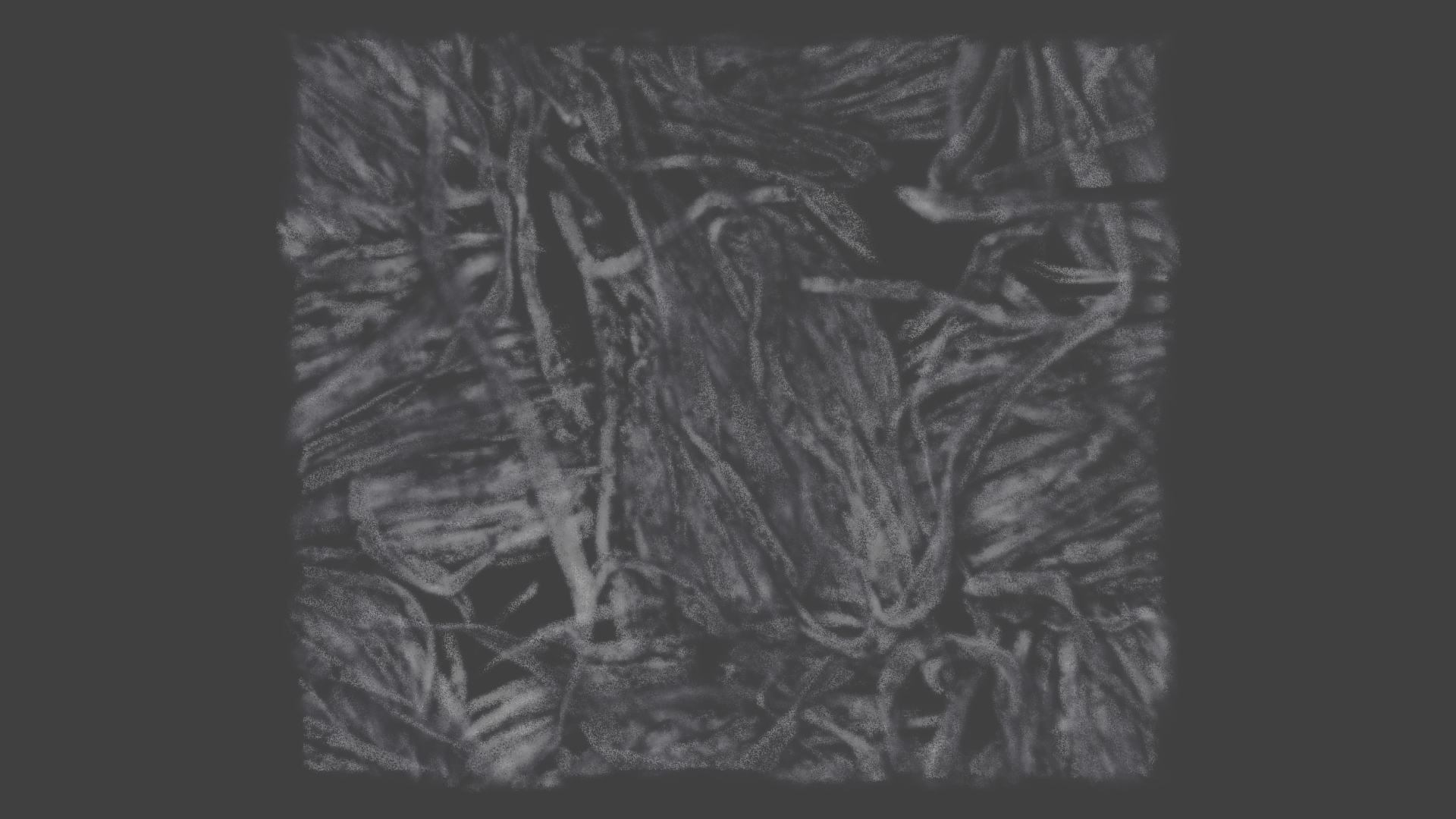}
      \caption{Snapshot of the movie in supporting information that shows the part of the fabric we calculate the flow field for. Rendering done using Blender \cite{blender}. (Multimedia view)}
      \label{fig:snapshot}
  \end{center}
\end{figure}

We want to model a representative part of the fabric of a face covering, so we study an area of two by two lattice sites. This area is shown by the white box in \Figref{fig:Ioatzin_image}, and in \Figref{fig:threshold+flow}(a). Note that we put the edges of the white rectangle in the densest part of the fabric where flow is least. The dimensions of the white rectangle are given in Table \ref{tab:fabric}. A full three-dimensional rendering of the region we study is shown in the Supporting Information, with a snapshot in \Figref{fig:snapshot}. \revs{The full image stack is available on Zenodo.}

\subsection{Estimation of what fraction of the fabric thickness is in our simulation box}

Using a mass density for cotton in Table \ref{tab:numbers}, then simply counting each voxel as
$(\SI{1.8}{\micro\metre})^3$ of cotton, we have a mass/unit area of cotton of $\SI{96}{\gram\per\metre\squared}$ in our fabric array of $330\times 280 \times 52$ voxels.
Our directly measured value is $\SI{120}{\gram\per\metre\squared}$, so this estimate is that our 52 slices or $\SI{93.6}{\micro\metre}$ of fabric contains 80\% of the mass of the fabric. However, our estimate for the fabric thickness using optical microscopy is $\SI{285}{\micro\metre}$, three times the thickness of our image.

\begin{figure}[htb!]
  \begin{center}
    \includegraphics[width=7.5cm]{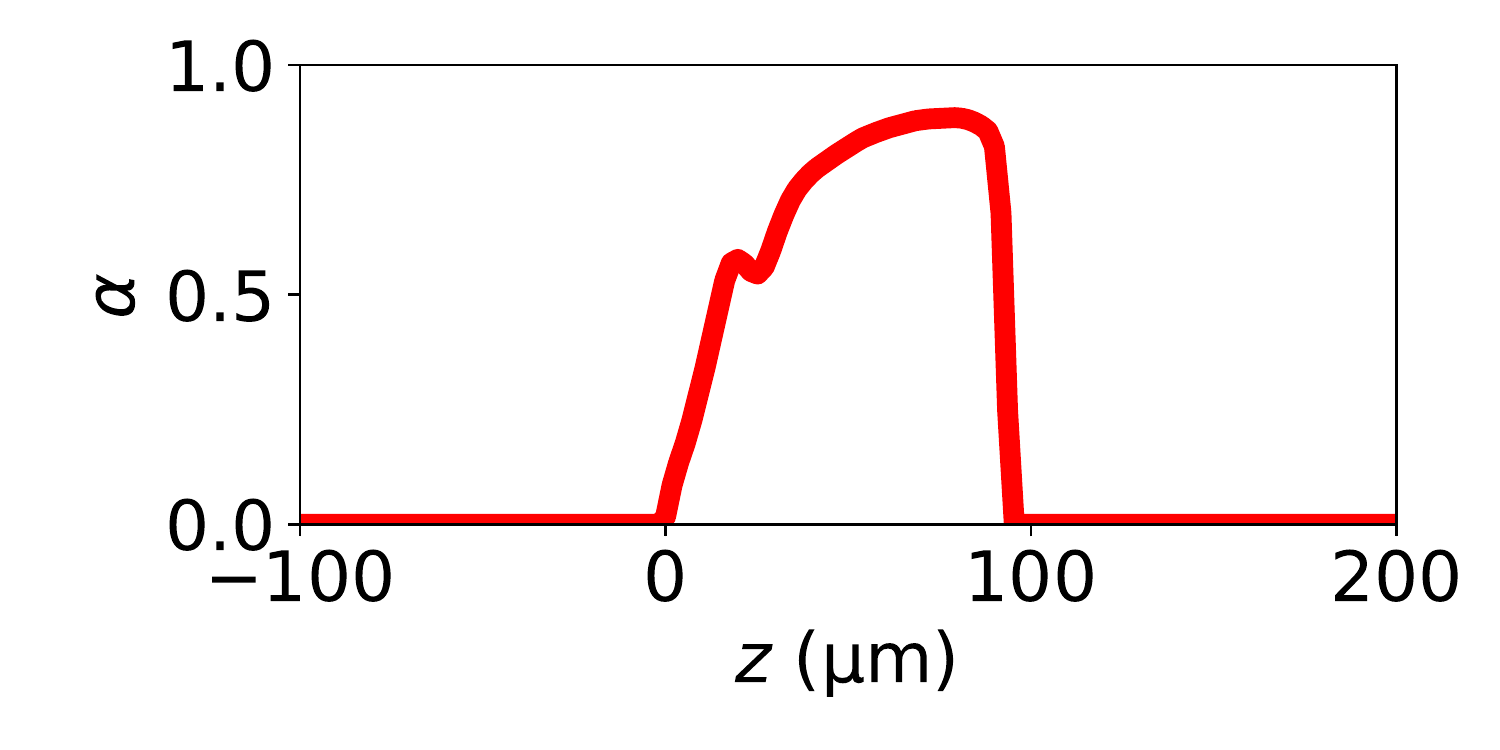}
      \caption{Plot of the fraction of voxels belonging to a fibre $\alpha$ (averaged over $x$ and $y$), as a function of $z$. The zero of $z$ is at the top of the fabric (slice 0). This is for the volume used in our simulations.}
      \label{fig:alphap}
  \end{center}
\end{figure}

The thickness of fabric measured in air is not perfectly well defined, the fabric is compressible being mostly air and at the edges there are stray fibres. We have plotted the average fraction $\alpha$ of voxels that are fibre voxels, as a function of $z$ in \Figref{fig:alphap}. Note that this is measured in solvent. It is mostly above the average value of 28\% we obtained in air, and the average value $\alpha$ inside the fabric of this plot is 69\%. It is possible that the fabric may have compacted and/or the fibres swollen in our solvent. 

To conclude, there is significant uncertainty in what fraction the fabric thickness is included in the 52 slices. We can only say that our 52 slices contains at least one third of the fabric, but probably no more than two-thirds.

\section{Lattice Boltzmann simulations of air flow through fabric}

Lattice Boltzmann (LB) simulations are performed on a three-dimensional lattice of $n_x$ by $n_y$ by $n_z$ lattice sites; $z$ is the flow direction.
Our code is the Palabos LB code from the University of Geneva \cite{palabos2020}. The code uses a standard one-relaxation-time LB algorithm on a cubic D3Q19 lattice. The speed of sound $c_s=1/\sqrt{3}$ in LB units where both the lattice spacing and the time step are set to one \cite{guo2013book}.
It has a kinematic viscosity
$\nuLB=c_s^2\left(\omega^{-1}-1/2\right)$. We set the relaxation rate $\omega=1$ in LB units, giving a kinematic viscosity $\nuLB=1/6$ in LB units \cite{guo2013book,behrend1994}.

We run the LB simulations until the change in mean flow speed along $z$ is very small so we are at steady-state.
We then insert particles into the resulting steady flow field to evaluate their trajectories.

Our code reads in the $330\times 280 \times 462$ array obtained from our image analysis. Fibre voxels have standard LB on-site bounce back \cite{ziegler1993,bao2011} to model stick boundary conditions \revs{for the air flow}. 

The box is configured such that the $x$ and $y$ edges are in denser parts of the fabric so there is little flow near and at these edges. 
In the LB simulations we use periodic boundary conditions (PBCs) along the $x$ and $y$ directions. The real fabric is not perfectly periodic and so our flow field has artifacts near the edges. However, there is no way of avoiding artifacts at the edges, and PBCs are a simple choice.


We impose a pressure gradient along the $z$ axis, to drive flow. We do this by fixing the densities in the first and last $xy$ slices of the lattice along $z$. We fix the density in the $z=0$ slice to be $1+10^{-5}$, and that in the $z=n_z-1$ slice to be $1-10^{-5}$. This corresponds to a pressure difference of $(2/3)\times 10^{-5}$ across the fabric. 

This small density/pressure difference across the fabric is chosen to keep the Reynolds number small, so we have Stokes flow.
The Reynolds number for flow with characteristic lengthscale $L$ is
\begin{equation}
\mbox{Re}=\frac{UL}{\nu}
\end{equation}
for $\nu$ the kinematic viscosity and $U$ the velocity. For the velocity we use the Darcy velocity, see section \ref{sec:Darcy}.
The Reynolds number for the largest lengthscale (yarn lattice constant along $x$) in our simulation box is in Table \ref{tab:fabric} and is much less than one so we have Stokes flow in our simulations.

For an air flow speed of $\SI{2.7}{\centi\metre\per\second}$ (moderate exercise) the Reynolds number for air flow with a characteristic lengthscale of a few hundred micrometres is $\Reynolds\simeq 1$. So in a fabric mask there will small deviations from Stokes flow, but we expect them to have little effect.

The LB simulations only give a flow field on a cubic lattice, so we use trilinear interpolation to give a continuous flow field $\vec{u}(\vec{r})$. Trilinear interpolation is the extension to three dimensions of linear interpolation in one dimension \cite{bourke_web}.

\begin{figure}[htb!]
  \begin{center}
    \includegraphics[width=7.5cm]{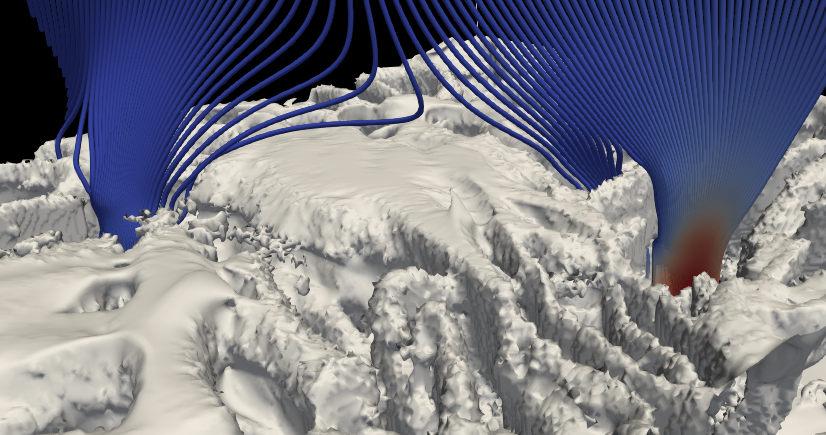}
      \caption{Plot of the fabric surface (white) together with streamlines. The streamlines are colour coded with local velocity: blue is slow, red is fast. The flat region in the centre of the image is the top of a yarn. Image produced by ParaView \cite{ahrens2005paraview}.}
      \label{fig:stream}
  \end{center}
\end{figure}

\section{Air flow through the woven fabric}

The air flow through the fabric is heavily concentrated in the inter-yarn pores, and there is essentially no flow through the centres of the yarns. This can be seen in the heatmap of the $z$ velocity in \Figref{fig:threshold+flow}(b).
Note that all the fastest voxels (shown in pale green) are in a single patch in the middle of the biggest inter-yarn gap. There are 718 of these voxels, out of 27,190 air voxels, and they contribute over a third of the total air flow through this slice.

The flow through the fabric is illustrated by streamlines in \Figref{fig:stream}. Note that all the streamlines shown flow around the yarns and through the gaps between the yarns. We conclude that as the air goes through inter-yarn pores, the filtration efficiency will depend on whether or not particles flowing through these pores, collide with the pore sides, or stray fibres across these pores.

The spacing between fibres of a yarn is mostly too small to be resolved by our imaging technique, so presumably is mostly a micrometre or less. Note that the integrity of yarns relies on large numbers of physical contacts \cite{warren2018}, so the fibres must touch in many places.
Our limited resolution means we cannot model any flow in between the fibres. However, as the inter-yarn gaps are $\sim\SI{50}{\micro\metre}$ across, the flow through any gaps between fibres of order $\sim\SI{1}{\micro\metre}$ or less will be negligible. \revs{Assuming that flow speeds through gaps scale as one over the gap size squared as it does in Poiseuille flow then any flow through the sub-micrometre inter-yarn gaps will be thousands of times slower than flow in the inter-yarn pores \cite{vandenbrekel1987,shin2018}.}

Finally, the fact that the bottom-right inter-yarn pores has the largest air flow illustrates that the fabric is disordered. It is not a perfect lattice of inter-yarn pores, each of which is the same. 
This also means that small (in the sense of difficult to detect with the naked eye) amounts of damage to fabric significantly affect flow through it.

\subsection{Darcy's law}
\label{sec:Darcy}


Fluid flow through fabric has been studied in earlier work on the washing of fabric (laundry). Removing dirt from fabric relies on the flow of water through the fabric \cite{vandenbrekel1987,shin2018,moholkar2004,bueno2018}. These earlier workers, starting with the pioneering work of van den Brekel \cite{vandenbrekel1987}, assumed that inter-yarn flow was dominant, which is corroborated by the present work. They modelled the flow through fabric using the standard approach for (low Reynolds number) flow through porous media: Darcy's Law.

A mask is a porous medium, and so at low Reynolds number the air flow $Q$ through the fabric is given by Darcy's Law \cite{whitaker1986}
\begin{equation}
Q=\frac{kA}{\mu}\frac{\Delta p_F}{L_F}
\label{eq:Darcy}
\end{equation}
which defines the permeability $k$. $Q$ is the volume of air crossing the fabric per unit time,
$A$ is the area of the fabric the air flows through and $\mu$ is the viscosity of air. 

For our thin fabric there are end effects.
We neglect these and just consider the pressure drop across the fabric, $\Delta p_F$ and the thickness of the fabric, $L_F$.
The flow $Q$ is proportional to the size of the pressure drop across the fabric $\Delta p_F$ and inversely proportional to the thickness $L_F$ of the fabric. The Darcy velocity $U$ is defined by
\begin{equation}
    U=\frac{Q}{A}
    \label{eq:udarcy}
\end{equation}
In free space $U$ is the actual flow velocity, while inside a porous medium, some of the area $A$ is occupied by the solid material and so does not contribute to $Q$. Then the local flow velocity varies from point to point and is mostly higher than the Darcy velocity $U$.

In our LB simulations we impose the pressure difference $\Delta p_F$ (via setting the densities at bottom and top along $z$), measure $Q$, and evaluate the permeability from
\begin{equation}
k=\frac{Q\mu}{A}\frac{L_F}{\Delta p_F}
\end{equation}
The viscosity of our LB fluid is $\mu=\rho_{LB}\nu_{LB}=1/6$, because $\rho_{LB}=1$ is the mass density in LB units and $\nu_{LB}=1/6$ is the kinematic viscosity also in LB units. In the same units $L_F=52$.

We find a permeability of $k\simeq 0.73$ in LB units, or  $k\simeq \SI{2.4}{\micro\metre\squared}$ on conversion using our known voxel size. This value is
comparable to the value $k\simeq \SI{4}{\micro\metre\squared}$ found for cotton sheets (with water as the fluid) in the experiments of van den Brekel \cite{vandenbrekel1987}.

Note that our fabric is imaged in liquid and van den Brekel's measurements are for fabric immersed in a liquid. So it is possible that in both cases the cotton may have swelled due to absorbing the liquid, reducing $k$. We imaged the masks in SEM (under vacuum) before and after immersion in tetralin for confocal imaging and observed no change. While of course it is possible that swelling occurred \emph{during} immersion in said solvent we find no evidence for irreversible change due to immersion in tetralin.

\subsection{Impedance and pressure drop across fabric}

The pressure drop across a mask must be low enough to allow easy breathing through the mask. As we have Stokes flow the pressure drop is linearly proportional to the flow velocity, and the proportionality constant defines the mask's impedance $I$ \cite{hancock2020} 
\begin{equation}
\Delta p_F=IU
\label{eq:impedance}
\end{equation}
Using \Eqref{eq:Darcy} and \Eqref{eq:udarcy}, we have
\begin{equation}
I=\mu L_F/k
\end{equation}
Using the viscosity of air and our estimated $k$, $I= \SI{7.1}{\pascal\second\per\centi\metre}$. This is the same order as Hancock \etal\cite{hancock2020} find for 300 TPI cotton. Konda \etal\cite{konda2020b} finds an impedance of $\SI{4.2}{\pascal\second\per\centi\metre}$ for a 180 TPI cotton/polyester blend. Sankhyan \etal\cite{sankhyan2021} find pressure drops in the range 40 to 55 Pa for a air speed of $\SI{8}{\centi\metre\per\second}$, which gives impedances in the range 5 to $\SI{7}{\pascal\second\per\centi\metre}$. 

Hancock et al.\cite{hancock2020} estimate that the American N95 standard for breathability requires a maximum impedance of around $\SI{30}{\pascal\second\per\centi\metre}$, four times our fabric's value. \revs{So we conclude that the impedance of our imaged fabric is well within the range of values that are easy to breath through.}


\subsubsection{Model for the Darcy's Law permeability}

Van den Brekel \cite{vandenbrekel1987} uses the Kozeny, or Kozeny-Carman, model for $k$. This model was developed for beds composed of packed spheres. Although as van den Brekel proposed the vast majority of the flow is through inter-yarn pores, these pores do not resemble the gaps between the sphere in beds of packed spheres. They are channels partially obstructed by stray fibres. Thus we model $k$ of our fabric by Poiseuille flow in cylinders of effective diameter $d_{EFF}$ that occupy an area fraction $\epsilon_{by}$ of the fabric. This gives
\begin{equation}
k\sim\frac{\epsilon_{by}d_{EFF}^2}{32}
\end{equation}
We estimate the effective free diameter to be in between a fibre diameter and a yarn diameter, $d_{EFF}\sim \SI{50}{\micro\metre}$, while the area fraction of inter-yarn pores $\epsilon_{by}\sim 0.1$. These values give $k\sim \SI{8}{\micro\metre\squared}$ --- the same order of magnitude as our measured value. \revs{Given the numerous approximations --- we estimate the channel size and pore fraction, the channels are too short for fully developed Poiseuille flow, and there are fibres that cross the channels --- we consider this reasonable agreement.} \revs{Bourrianne \etal\cite{bourrianne2021} find a similar value, $k= \SI{12}{\micro\metre\squared}$ for a surgical mask.} This is consistent with the flow being predominantly through pores tens of micrometres across, that occupy about ten percent of the total area.

\subsection{Curvature of streamlines}

The inertia of a particle only affects its motion when streamlines are curving. For flow that is just straight ahead the particle will just follow the flow. So we need to characterise the curvature of the streamlines going through the fabric. We do this by determining a characteristic lengthscale for this curvature, which we call $\Sigma$.

The lengthscale $\Sigma$ for curvature of a streamline at a point on a streamline of the flow field is defined by
\begin{equation}
    \Sigma = \frac{{\vec u}.{\vec u}}{a_{\bot}}
\end{equation}
for ${\vec u}$ the flow field at that point, and $a_{\bot}$ the magnitude of the normal component of the acceleration ${\vec a}$ along the streamline at this point. Streamlines are defined by velocities and accelerations and so one way to obtain a lengthscale is the square of a velocity divided by an acceleration.

The acceleration is that along the streamline, i.e., rate of change of streamline velocity while being advected along the streamline. The normal component is obtained by subtracting the parallel component, from ${\vec a}$
\begin{equation}
    {\vec a_{\bot}}=a-\hat{u}(\hat{u}.\vec{a})
\end{equation}

\begin{figure}[htb!]
  \begin{center}
    \includegraphics[width=7.5cm]{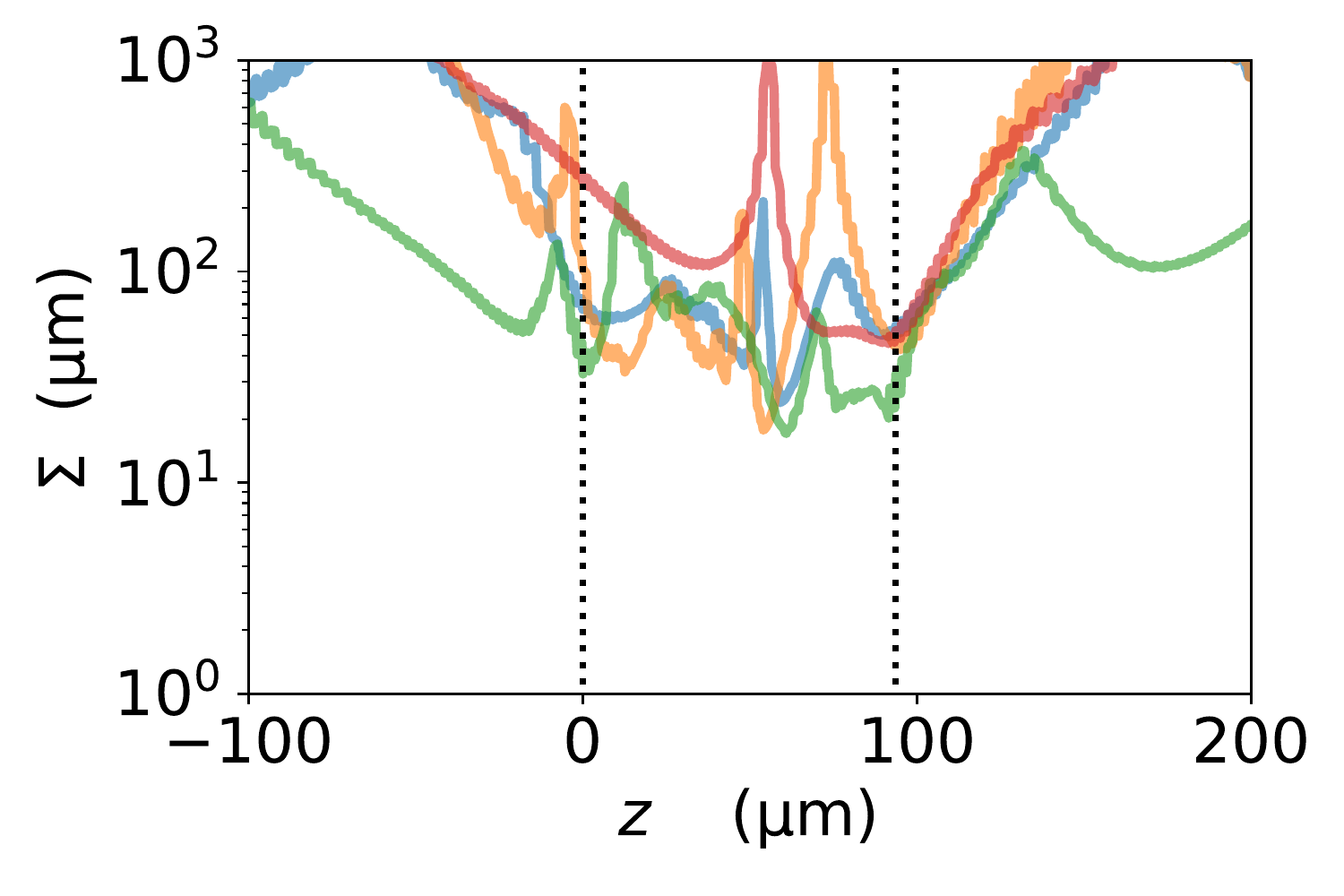}
      \caption{Plot of the local curvature $\Sigma$ along four streamlines, as a function of their position along the flow direction $z$. The vertical dotted lines mark the start and end of the fabric, so outside of these lines we are outside the fabric. N.B. the curves are not smooth because $\Sigma$ depends on an acceleration. The flow field velocity is obtained by interpolation so the velocity is continuous but its derivative the acceleration is not.}
      \label{fig:Sigma}
  \end{center}
\end{figure}

We have plotted $\Sigma$ along a set of streamlines in \Figref{fig:Sigma}. The local curvature along streamlines within the fabric varies greatly but is mostly around tens to hundreds of micrometres. This is different from the flow in a mesh of single fibres, as found in surgical masks. In surgical masks there is only one lengthscale, that of the fibre diameter, which is typically around $\SI{15}{\micro\metre}$ \cite{robinson2021}. So in non-woven filters such as surgical masks, the curvature lengthscale is expected to approach $\SI{15}{\micro\metre}$ for trajectories near the surfaces of fibres.

\section{Calculating particle trajectories and collisions}

In this section we first introduce the theory for particles moving in a flowing fluid, then describe the details of our calculations.

\subsection{Theory for a particle in a flowing fluid}

The particles are spheres of diameter $d_p$, that feel only the Stokes drag of the surrounding air. We neglect any perturbation by the particles of the flow field, and assume that the drag force on a particle couples to its centre of mass. Then Newton's Second Law for the particle becomes
\begin{equation}\label{eq:particle-newton2}
  m_p \frac{{\rm d} \vec{v}}{{\rm d} t} = - 
  \frac{3\pi\mu d_p}{C}
  \left(\vec{v} - \vec{u}\right) 
\end{equation}
for a particle of mass $m_p$ and velocity $\vec{v}$ in a flow field $\vec{u}$ of fluid with viscosity $\mu$. Here $C$ is the Cunningham slip correction factor \cite{lee1980,kanaoka1987}. We consider particles with $d_p\geq\SI{1}{\micro\metre}$ (due to limited imaging resolution). In this size range $C$ is always close to one (within 15\%). Therefore, we just set $C=1$ here.

The particles are spheres of mucus which we assume has the mass density of water, $\rho_p$. Then $m_p=(\pi/6)d_p^3\rho_p$, and \Eqref{eq:particle-newton2} becomes
\begin{equation}\label{eq:particle-newton3}
  \frac{{\rm d} \vec{v}}{{\rm d} t} = - 
  \frac{18\mu}{\rho_pd_p^2C}
  \left(\vec{v} - \vec{u}\right) =
  -\frac{\left(\vec{v} - \vec{u}\right)}{t_I}
\end{equation}
where we have introduced $t_I=\rho_pd_p^2C/(18\mu)$: the timescale for viscous drag to accelerate the particle.

\subsubsection{The Stokes number}

\revs{When integrating \Eqref{eq:particle-newton3} if the timescale $t_I$ is short then the particle closely follows (the streamlines of) the fluid flow, and so when the fluid flows round an obstacle, the particle follows the fluid. However, if $t_I$ is large then when the fluid flow changes direction the particle's inertia results in it carrying on moving in the direction of the fluid before it changed direction. This inertial effect can result in a particle colliding with an obstacle, although the fluid flows round it, and is the cause of inertial filtration \cite{wang2013}. Short and long timescales $t_I$ are relative to the timescale for the change of direction of the fluid flow, and the ratio of these two timescales defines a dimensionless number: the Stokes number.}

The ratio of the timescale $t_I$ to the timescale for fluid flow to change direction as it goes round an obstacle of size $L_O$, defines the Stokes number
\begin{equation}\label{eq:stokes-number1}
  \St(d_p,L_O,U)
  = \frac{t_I}{L_O/U}
\end{equation}
where we use the Darcy speed $U$. Then
\begin{equation}\label{eq:stokes-number}
  \St(d_p,L_O,U)
  = \frac{ \rho_p d_p^2 U C}{18\mu L_O}
  \sim \frac{\num{3.08e6}}{\si{\metre\squared\per\second}} \frac{d_p^2}{L_O} U ,
\end{equation}
Parameter values in table \ref{tab:numbers} were used.
For $\St\ll 1$ viscous forces dominates inertia and the particle follows streamlines faithfully. However, for $\St\gg 1$, inertia dominates and the particle's trajectory will strongly deviate from streamlines. As the streamlines go round obstacles, deviating from streamlines can result in the particle colliding with an obstacle and being filtered out. This is inertial filtration.

\begin{figure}[htb!]
  \begin{center}
    \includegraphics[width=7.5cm]{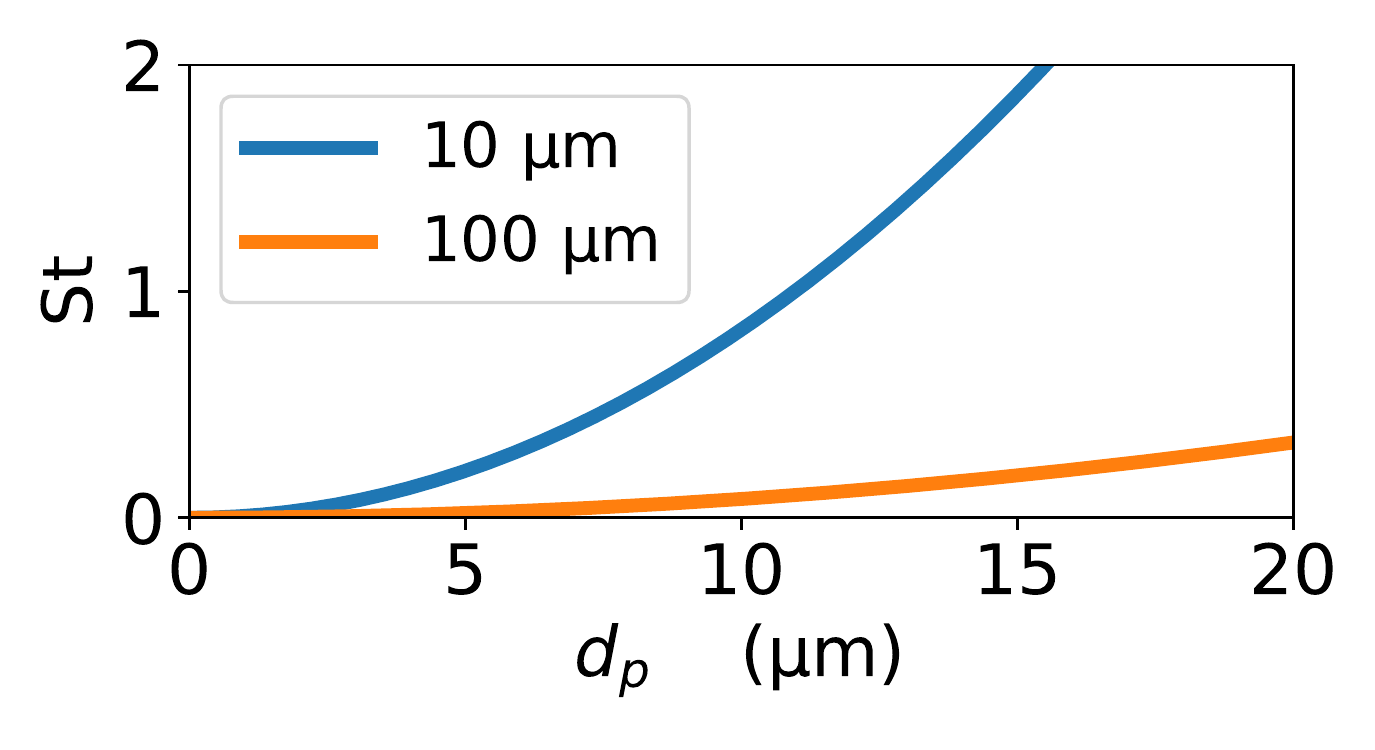}
      \caption{Plot of the Stokes number as a function of particle diameter $d_p$, using \Eqref{eq:stokes-number}. The blue and orange curves are for obstacle sizes $L_O=\SI{10}{\micro\metre}$ and $\SI{100}{\micro\metre}$ respectively. The flow speed is set to $U=\SI{2.7}{\centi\metre\per\second}$.}
      \label{fig:Stokes}
  \end{center}
\end{figure}

The Stokes number depends on the flow speed, and on both the size of the particle and of the obstacle the flow is going around.
Figure \ref{fig:Stokes} shows the Stokes number as a function of particle diameter, for particles in flows fields curving over lengthscales of $10$ and $\SI{100}{\micro\metre}$. Note that for flow fields curving over a distance $\SI{10}{\micro\metre}$, a Stokes number of one is only reached for particles greater than $\SI{10}{\micro\metre}$ in diameter. So our fabric where the curvature $\Sigma$ is mainly at least tens of micrometres (see \Figref{fig:Sigma}) is expected to show little inertial filtration of any particle around $\SI{10}{\micro\metre}$ or smaller in diameter.

\subsection{Evaluation of filtration using our Lattice-Boltzmann flow field}

The filtration efficiency is estimated from the fraction of the particles that collide with the fabric.
\revs{We calculate the trajectories of $N_{samp}$ particles that start in a uniform grid that occupies the central quarter of the area in the white rectangle in \Figref{fig:Ioatzin_image}. This area in the white rectangle is two lattice constants of the fabric across along both the $x$ and $y$ axes, and so the area the particles start from fills one unit cell of the fabric lattice. Our filtration efficiency should therefore  be a good representation of the average filtration efficiency of a large area of fabric. Once we have computed the trajectories of the $N_{samp}$ particles and determined which ones collide with the fabric, the filtration efficiency is computed from}
\begin{equation}
    \mbox{Filtration efficiency}=\frac{\sum_i^{\mbox{coll}}v_{zi}}{\sum_i^{\mbox{coll}}v_{zi}+\sum_i^{\mbox{pen}}v_{zi}}
    \label{eq:filterflux}
\end{equation}
where the sum with superscript \lq coll\rq~is over all particles that collided with a fibre voxel, and the sum with superscript \lq pen\rq~is over all particles that pass through the fabric without colliding. $v_{zi}$ is the $z$ component of the  velocity of particle $i$ at the starting point of its trajectory. Note that as we are interested in the fraction of the \revs{particles} filtered, each particle is weighted by the local velocity. We assume the particle concentration is uniform in the air, so regions where the air is flowing faster contribute more than where the regions are flowing more slowly. 

See Appendix \ref{app:traj} for further details of how we compute trajectories, \revs{and the condition for collisions}.
All calculations are for flow at the speed $U=\SI{2.7}{\centi\metre\per\second}$, corresponding to breathing under moderate exertion (see Table \ref{tab:numbers}).


\begin{figure}[htb!]
  \begin{center}
    \includegraphics[width=7.5cm]{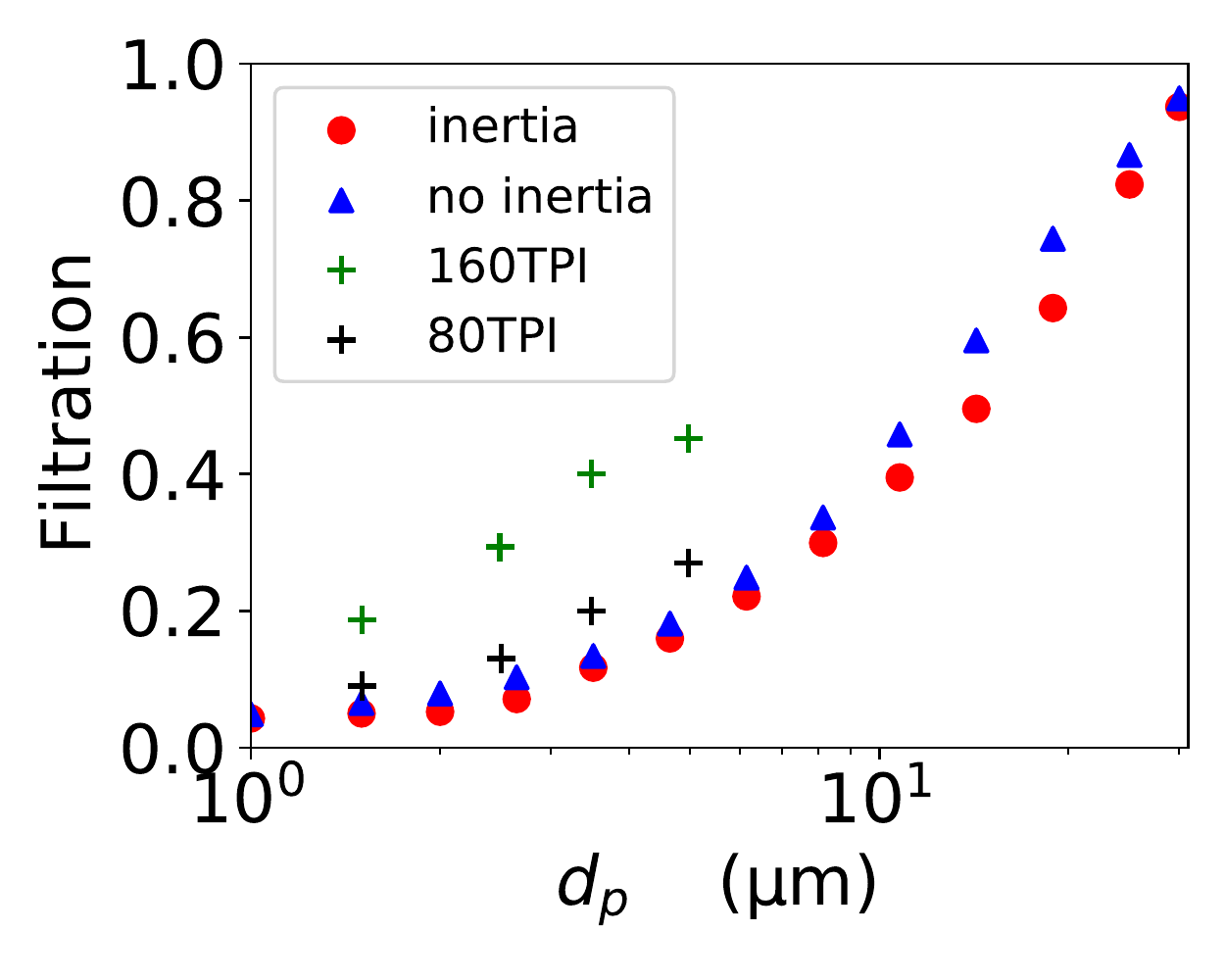}
      \caption{Plot of the fraction of particles filtered, as a function of their diameter $d_p$. This is in air with flow speed $U=\SI{2.7}{\centi\metre\per\second}$. The red circles \revs{are with the inertia of a particle with the mass density of water and the blue triangles are without inertia}. They
      are each averages over $N_\mathrm{samp}=1600$ particle trajectories. The green and black pluses are measurements of Konda et al.\cite{konda2020b} (obtained from Fig.~2(B)\cite{Rohatgi2020}). These measurements are for a pressure drop across the fabric of $\SI{10}{\pascal}$, whereas at our value of $U$, the estimated pressure drop is $\SI{19}{\pascal}$. The impedances measured by Konda et al.\cite{konda2020b} are lower than our value ($\SI{7.1}{\pascal\second\per\centi\metre}$), they find values of $\SI{1.3}{\pascal\second\per\centi\metre}$ for 80 TPI, and $\SI{4.2}{\pascal\second\per\centi\metre}$ for 160 TPI. Thus, especially for the 80 TPI fabric, although their pressure drop is lower, the air speed is higher.}
      \label{fig:filtration}
  \end{center}
\end{figure}

\section{Results for particle filtration}

In \Figref{fig:filtration} we have plotted results for the fraction of particles that collide with a fibre and are filtered out, as a function of the diameter of the particle. These are the red data points. We see that the efficiency is less than 10\% for micrometre sized particles, and although it increases with increasing size we are still filtering less than half of the particles at a diameter of $\SI{10}{\micro\metre}$. We breathe out droplets with a wide range of sizes but the peak of this distribution is around one micrometre \cite{johnson2011}. We predict that the fabric we have imaged is very poor at filtering out droplets of this size. But note that we could only image approximately half of one cotton fabric layer; presumably the filtration efficiency of the full layer is higher.

Both Konda \etal\cite{konda2020,konda2020b} and Duncan \etal\cite{duncan2020} have measured the filtration efficiency of woven fabrics, for particles up to five micrometres. Both groups find a large variability in filtration efficiency from one material to another, with filtration efficiencies in the range less than 10 to almost 100\%, for particles with diameters of a few micrometres. Sankhyan \etal\cite{sankhyan2021} found comparable filtration efficiencies to Konda \etal and Duncan \etal. They also found that the fabric masks were systematically less good at filtering than non-woven surgical masks.

Two data sets from Konda et al.\cite{konda2020b} are plotted in \Figref{fig:filtration}.
Konda et al.\cite{konda2020,konda2020b} found that the filtration efficiency of fabric increased with its TPI. In \Figref{fig:filtration} we see that they found that the filtration efficiency for 160 TPI cotton/polyester fabric is higher than for 80 TPI cotton. We estimate that our fabric's TPI is 186. Our efficiencies are lower than those measured by Konda et al.\cite{konda2020b} but the slope is very similar. At a diameter of $\SI{1.5}{\micro\metre}$ we find an efficiency of 5\%, whereas Konda et al.\cite{konda2020b} find efficiencies of 9 and 19 \% for TPIs of 80 and 180, respectively. Our model makes a number of approximations: flow field on a $\SI{1.8}{\micro\metre}$ lattice, \revs{possible changes in the fibres and yarns due to immersion in the solvent}, coupling at centre of mass, etc, so our estimated efficiencies are likely only accurate to within a factor of two \revs{in either direction. With this estimate of the uncertainty in our calculation, we estimate an efficiency in the range 2.5 to 10\%.} Thus, within our large uncertainties our results are essentially consistent with the measurements.


\subsection{Inertia can cause collisions to be avoided and so reduce filtration efficiency}

In order to understand the role of inertia in filtration by woven fabric, we calculated the filtration without inertia. \revs{In other words the Stokes number is zero and the particles follow the streamlines perfectly.} The results are shown as blue \revs{triangles} in \Figref{fig:filtration}, and are for pure interception filtration. If we compare those points with the red points, which are with inertia, we see that the difference is small. Inertia has a small effect and filtration is mainly interception.

\begin{figure}[htb!]
  \begin{center}
    \includegraphics[width=7.5cm]{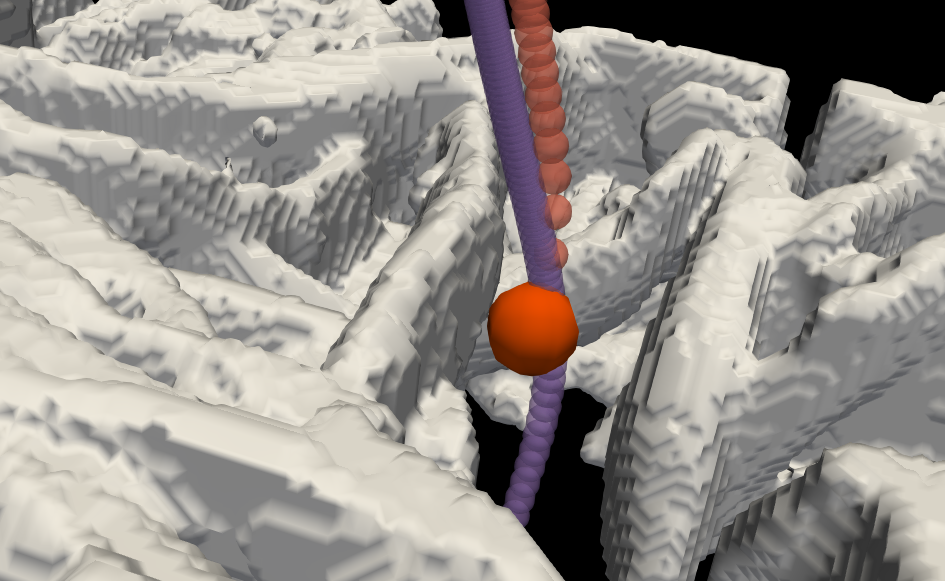}
      \caption{A pair of trajectories with and without inertia, that start at the same point. This is for a particle of diameter \SI{20}{\micro\metre}. The fabric is shown in white, and trajectories with and without inertia are traced out by purple and by orange spheres, respectively. The sphere at the collision point is shown at the true particle size, others along the path are smaller, for clarity. Note that with inertia the particle penetrates the fabric while without it, the particle collides at the point shown by the large orange sphere. Here inertia carries the particle a little farther out from the side of the inter-yarn pore, avoiding a collision.  Image produced with ParaView \cite{ahrens2005paraview}.}
      \label{fig:inertialtraj}
  \end{center}
\end{figure}

But the difference is that the effect of inertia is to slightly decrease filtration. We have found that the effect of inertia can be to cause a collision that occurs without inertia to be avoided, see \Figref{fig:inertialtraj}. There we have plotted two trajectories with the same starting point but with inertia (purple) and without inertia (orange). The particle with inertia penetrates the fabric, while without inertia it collides with the side of the inter-yarn pore and is filtered out. Inertia carries a particle closer to the centre of an inter-yarn pore where it is further from the sides and so escapes colliding with these walls.

\revs{In the standard picture of filtration of particles from air, the effect of inertia is always to increase filtration efficiency \cite{wang2013}. In that standard picture, deviations of particle trajectories from streamlines due to inertia always increase the probability of a collision. This is not we have found, see \Figref{fig:inertialtraj}. Here} and in Robinson \etal\cite{robinson2021} we find that at small Stokes numbers the situation can be more complex and subtle. \revs{Inertia at small Stokes number} can make filtration a little less efficient.  \revs{However,} at large Stokes number we indeed find that inertia increases filtration efficiency.


\revs{The zero Stokes number (i.e., zero inertia) limit, often called interception filtration \cite{wang2013}, corresponds to the limit in which the air speed $U\to 0$ --- as $U=0$ gives a Stokes number of zero. Thus, we have shown that reducing $U$ from a speed characteristic of moderate exercise to zero, has little effect on the filtration efficiency.  Filtration} by our fabric is almost independent of $U$, or equivalently of the pressure drop across the fabric. This is in agreement with findings of Konda \etal\cite{konda2020b} who found that filtration did not vary significantly when they varied the pressure drop across the sample.

\section{Filtration via particles diffusing into contact}
\label{sec:diffusion}

Filtration of particles of order $\SI{100}{\nano\metre}$ is typically dominated by diffusion of the particles onto the surfaces of the filter \cite{wang2013}. The nanoparticles then stick and are filtered out. With a flow field based on imaging at $\SI{1.8}{\micro\metre}$ resolution, we are unable to be quantitative about the filtration efficiency for particles in this size range. But we are able to argue that the efficiency of filtration by diffusion should be low. The argument is as follows.

For our fabric, almost all air flows through inter-yarn pores $\sim\SI{50}{\micro\metre}$ across. So filtration by diffusion relies on a particle diffusing across the flowing air stream into contact with the sides of the inter-yarn pore, during the short time the particle is being advected through the fabric. So filtration efficiency is determined by the ratio of a diffusive time $t_{DX}$, to an advection time $t_A$. $t_{DX}$ is the time taken to diffuse across (i.e., in $xy$ plane) an inter-yarn pore. $t_A$ is the time taken for air to flow through the pore.

The ratios of diffusive to flow timescales are called P\'{e}clet numbers. Here the P\'{e}clet number is
\begin{equation}
    {\rm Pe}=\frac{t_{DX}}{t_{A}}
    \label{eq:peclet1}
\end{equation}
For a particle $\SI{100}{\nano\metre}$ in diameter, Stokes-Einstein gives $D=kT/(3\pi\mu d_p)\sim \SI{240}{\micro\metre\squared\per\second}$, and so for a distance of $\SI{50}{\micro\metre}$, $t_{DX}\sim (50^2)/80\sim \SI{10}{\second}$.
The advection timescale is just the time taken for air to flow through the fabric $t_{A}\sim \SI{100}{\micro\metre}/\SI{2.7}{\centi\metre\per\second}\sim\SI{4}{\milli\second}$. Thus
\begin{equation}
    {\rm Pe}\sim 3000
    \label{eq:peclet2}
\end{equation}
As ${\rm Pe}\gg 1$ then particles with $d_p=\SI{100}{\nano\metre}$ are carried through the fabric much faster than they can diffuse across the inter-yarn pores, and we expect the efficiency of filtration by diffusion to be very low. Note that for larger particles $D$ is smaller so filtration by diffusion is even less efficient.

\revs{Our prediction that filtration via diffusion should be very inefficient is consistent with a number of experimental studies \cite{konda2020b,duncan2020,drewnick21,zangmeister2020}. These studies all find that woven fabrics are poor at filtering particles much less than a micrometre in diameter, which is the size range where particle diffusion is fastest. Here poor filtration means typically less than 50\%, and in some cases much less. For diameters less than a micrometre, woven fabrics are typically poorer filters than the non-woven materials used in surgical masks. For the non-woven materials in surgical masks, at diameters around $\SI{100}{\nano\metre}$, the efficiency increases as the diameter decreases, due to diffusion becoming increasingly important as the diameter increases \cite{wang2013,robinson2021,robinsonPanic2020,konda2020b,duncan2020,drewnick21,zangmeister2020}. This increase is also seen in woven fabrics \cite{konda2020b,duncan2020,drewnick21,zangmeister2020} but is mostly weaker for woven than for non-woven fabrics.}

\section{Conclusion}

\revs{Measurements of the filtration efficiency of woven fabrics consistently find poorer filtration efficiency than for the non-woven materials used in surgical masks or other air filters \cite{konda2020b,zangmeister2020,duncan2020,drewnick21,hancock2020}. This is for the filtration of particles both smaller than and larger than a micrometre, and for a range of different fabrics of different TPIs and materials (cotton, polyester, etc). For the first time}, we have the complete flow field (at a resolution of $\SI{1.8}{\micro\metre}$) inside the fabric, and we can also control the inertia of the particles, so we can see why the efficiency is so low. The efficiency is low because essentially all the air flows through relatively large (tens of micrometres) inter-yarn pores, which are only obstructed by a few stray fibres, see \Figref{fig:snapshot}.  Particles just follow the air through these gaps and so few are filtered out. 

\revs{Inter-yarn pores will vary in size from one woven fabric to another, for example they should be smaller when the TPI is larger. Some data suggests that
fabrics with higher TPIs are better filters \cite{konda2020b}, possibly because the inter-yarn pores are smaller. However, all woven fabrics are made of yarn and so all will have inter-yarn pores. This together with the multiple experimental studies finding poor filtration efficiency \cite{konda2020b,zangmeister2020,duncan2020,drewnick21,hancock2020}, suggests that poor filtration is generic, because as we have seen particles are just carried through the relatively large inter-yarn gaps. These gaps are an order of magnitude greater in size than typical fibre spacings in the non-woven material in surgical masks \cite{robinson2021,lee2021b}.}

We estimate that the filtration efficiency of our imaged fabric is \revs{in the range 2.5 to 10\%.} This is for particles of diameter $\SI{1.5}{\micro\metre}$, which is around the most probable size for droplets exhaled while speaking \cite{johnson2011}. \revs{Thus this is the most probable droplet size for source control. When we consider protection of the mask wearer, we consider inhalation of droplets from the surrounding air. Then we need to consider droplets that have evaporated in the surrounding air.} \revs{Because the filtration efficiency decreases with decreasing particle size, it} will be even lower for droplets once they have \cite{netz2020,johnson2011,robinsonPanic2020} entered room air, and evaporation has \revs{reduced} their diameter \revs{by a factor of two to three \cite{johnson2011,robinsonPanic2020}.}  Our filtration efficiency is for approximately half a layer of woven fabric with an estimated TPI of 186. Konda \etal\cite{konda2020b} found filtration efficiencies of 9\% and 18\% for (complete single layers of) woven fabrics of 80 and 160 TPI. Sankhyan \etal\cite{sankhyan2021} found similar values.

\subsection{\revs{It may be impossible to make good filters from woven fabrics}}

Filtration \revs{by fabric} can be improved by using multiple layers \cite{sankhyan2021}. However, both multiple layers and higher TPI lead to higher impedance to air flow. 
\revs{Making a practical air filter always involves a trade off between maximising filtration, and keeping the impedance (pressure drop) low enough to be acceptable to the user. In other words, the 
\begin{equation}
\mbox{figure of merit for a filter}=\frac{-\ln[1-\mbox{Fraction Filtered}]}{I}
\end{equation}
}

\revs{Our estimated impedance of $I=\SI{7.1}{\pascal\second\per\centi\metre}$ is low in the sense that it is approximately one quarter the maximum impedance allowed by the Ameican N95 standard \cite{hancock2020}. However, due to the very low filtration efficiency, the value of the figure of merit is low for our fabric. Taking our 5\% filtration efficiency for $\SI{1.5}{\micro\metre}$, our estimated figure of merit is $\SI{0.007}{\centi\metre\per\pascal\per\second}$. Achieving 95\% filtration at the maximum impedance allowed by an N95 mask requires a figure of merit of $\SI{0.1}{\centi\metre\per\pascal\per\second}$, more than ten times the value for our cotton fabric. Here we used
Hancock \etal's\cite{hancock2020} estimated maximum impendance of the N95 standard of $\SI{30}{\pascal\second\per\centi\metre}$. It may be that it is impossible or almost impossible to make good filters from fabrics, because their figures of merit for filtration are too low.}

\subsection{\revs{The effect of particle inertia on filtration}}

\revs{We find that for our woven fabric,}
filtration is \revs{mostly} due to interception over the size range from one to a few tens of micrometres. \revs{In other words, filtration is due to particles that largely follow the streamlines but collide with cotton fibres due to the particle's size \cite{wang2013}. Note that filtration is only weakly affected by setting the inertia of particles to zero, compare the blue and red points in \Figref{fig:filtration}.} Surprisingly, over this size range, the effect of inertia is to decrease filtration efficiencies, although the effect is small. Modest amounts of inertia decrease filtration efficiency by pushing more particle trajectories away from collisions with fibres, than they do trajectories towards collisions. Very large amounts of inertia (for example due to a sneeze greatly increasing $U$) will increase efficiency due to most of the fabric area being occupied by yarns.

The non-woven filters in surgical masks and respirators (such as the European standard FFP and American standard N95 respirators), \revs{ force the air around single fibres of typical size around $\SI{5}{\micro\metre}$\cite{lee2021b}}. This smaller lengthscale \revs{for the curvature of streamlines} in surgical masks brings inertial filtration into play for droplets around a few micrometres in diameter \cite{robinson2021}. \revs{This makes inertial filtration much more effective for surgical masks and respirators than for woven fabrics, for particles one or a few micrometres in diameter.}





\subsection{Limitations of the present work, and future work}

We have simulated the flow field through one sample of woven fabric at a resolution of $\SI{1.8}{\micro\metre}$, and used this to understand the observed poor filtration performance. Future work could look at different fabrics, with different TPIs, and go to higher resolution, \revs{as well as compare with the materials used in surgical masks \cite{lee2021,du2021}.} Higher resolution images will improve the estimation of the filtration of smaller particles in particular, as this is likely to be sensitive to yarn/fibre roughness of lengthscales of a micrometre and smaller.


\begin{acknowledgments}
The authors with to thank Fergus Moore, Jonathan Reid and Patrick Warren for very useful discussions. We also acknowledge the support of the University of Surrey's High Performance Computing centre. Finally the authors would like to thank Judith Mantell from the Wolfson Bioimaging Facility (EPSRC Grant “Atoms to Applications”, EP/K035746/1) and Jean-Charles Eloi of the Chemical Imaging Facility, for the SEM images and assistance in this work.

\revs{RPS has undertaken consultancy for PolarSeal Tapes and Conversions Ltd, Farnham, UK.}
\end{acknowledgments}

\section*{Data Availability Statement}

The data and computer code that support the findings of this study are openly available in Zenodo at \url{http://doi.org/10.5281/zenodo.5552357}.

\appendix

\section{Computational details for integrating particle trajectories}
\label{app:traj}

Each particle trajectory is obtained by starting the particle at $z=5$ in LB units, and at $x$ and $y$ coordinates on a square grid in the central quarter of the box, i.e., from $n_x/4$ to $3n_x/4$ along the $x$ axis and from $n_y/4$ to $3n_y/4$ along the $y$ axis. We varied the starting region for the trajectories to observe the dependence of efficiency on starting region, and the efficiency varied by amounts around 10\%. The particle starts with the same velocity as the local flow velocity. Weighting the trajectories by their initial velocities using \Eqref{eq:filterflux} makes a difference of approximately twenty percent for our box with 200 LB lattice spacings in front of the fabric. It makes more of a difference for shorter boxes along $z$, hence our box size is trade off between accuracy and computational cost. 

Of order 20\% of trajectories leave the box at the sides. These are are not counted in our flux calculations. Although the LB flow field has periodic boundary conditions at the sides, this does not reproduce well the true conditions in the fabric, which is not perfectly periodic in $x$ and $y$. We cannot use a larger box along $x$ and $y$ because the larger box reaches the edge of the fabric strip, and a defect in the fabric, see \Figref{fig:Ioatzin_image}.

So we have multiple sources of uncertainties, each ten or a few tens of percent. Plus we only couple the particle to the fluid flow at the particle's centre of mass, and are using a flow field with spatial resolution larger than the smallest particles we consider. Considering all these sources of uncertainty, and the approximations of the model, we estimate that our results are accurate to about a factor of two. 

Each trajectory is integrated forward in time, using adaptive-step-size modified Euler integration of \Eqref{eq:particle-newton3}, until the particle either collides with a fibre voxel, or reaches the bottom (large $z$) edge of the simulation box. At each time step, we check for a collision.
A collision occurs if the centre of the particle is within a distance $(1/2)(d_p+\delta)$, i.e., radius of the particle plus a correction $\delta$. We estimate that the optimal value of $\delta$ is $0.5$ in LB units. So we use this value throughout this work.

The integration of \Eqref{eq:particle-newton3} requires we determine $t_I$ in LB units. This is done as follows, for the example of a particle with $d_p=\SI{5}{\micro\metre}$. First, we obtain the mean velocity in the LB flow field in a slice far from the fabric, as $U=5.8\times 10^{-7}$ in LB units. Second, we use \Eqref{eq:stokes-number} to determine that $\St=1.16$, for lengthscale $L=\SI{1.8}{\micro\metre}$ and $U=\SI{2.7}{\centi\metre\per\second}$. Then we use \Eqref{eq:stokes-number1} in LB units to obtain, with $L=1$ and our LB $U$, that $t_I=20.8\times 10^6$ in LB units. This value of $t_I$ reproduces the correct Stokes number for a particle $\SI{5}{\micro\metre}$ in diameter. The particle then collides with any lattice site within a distance of $1.89$ LB units.



%

\end{document}